\documentclass[10pt,conference]{IEEEtran}
\usepackage{cite}
\usepackage{amsmath,amssymb,amsfonts}
\usepackage{algorithmic}
\usepackage[ruled,vlined]{algorithm2e}
\usepackage{graphicx}
\usepackage{textcomp}
\usepackage{xcolor}
\usepackage{float}
\usepackage[dvipsnames]{xcolor}
\usepackage{mathtools}
\usepackage{multicol}
\usepackage{booktabs}
\usepackage{changes}
\usepackage{braket}
\usepackage{hyperref}
\definechangesauthor[name={kien}, color=red]{kn}
\definechangesauthor[name={ilya}, color=blue]{is}

\def\BibTeX{{\rm B\kern-.05em{\sc i\kern-.025em b}\kern-.08em
    T\kern-.1667em\lower.7ex\hbox{E}\kern-.125emX}}
\begin{document}

\title{Graph-Conditioned Meta-Optimizer for QAOA Parameter Generation on Multiple Problem Classes
}

\author{
\IEEEauthorblockN{
Kien X. Nguyen\IEEEauthorrefmark{1},
Ilya Safro\IEEEauthorrefmark{1}\IEEEauthorrefmark{2}
}
\IEEEauthorblockA{\IEEEauthorrefmark{1}\textit{Department of Computer and Information Sciences, University of Delaware}, Newark, DE, USA}
\IEEEauthorblockA{\IEEEauthorrefmark{2}\textit{Department of Physics and Astronomy, University of Delaware}, Newark, DE, USA}
\IEEEauthorblockA{\{kxnguyen, isafro\}@udel.edu}
}

\maketitle

\begin{abstract}
We study parameter transferability for the Quantum Approximate Optimization Algorithm (QAOA) across multiple combinatorial optimization problem classes from a parameter generation perspective. 
Specifically, a meta-optimizer is trained on one problem class and deployed on another during test time.
Prior work employs a Long Short-Term Memory network to emulate QAOA optimization trajectories, but the learned dynamics usually collapse to near-identical paths, limiting cross-problem transfer efficiency.
In this paper, we present a \textit{problem-aware graph-conditioned} meta-optimizer for QAOA that learns to generate parameter trajectories over a fixed horizon, providing strong initializations with only a few steps. 
The optimizer is conditioned on compact graph embeddings and trained end-to-end using differentiable feedback from the QAOA objective, avoiding the need for ground-truth angles.
We evaluate across multiple graph problem classes, including MaxCut, Maximum Independent Set, Maximum Clique, and Minimum Vertex Cover.
We report both solution quality and feasibility-aware metrics where constraints apply. 
Results across a comprehensive empirical study consisting of 64 settings show that the learned optimizer can reduce optimization effort and improve performance over standard initialization, while exhibiting transferable behavior across graph families and problem types.\\
{\bf Reproducibility: } Our code is available at: \url{https://github.com/Nyquixt/Cross-Problem-QAOA-ParamGen}.
\end{abstract}


\section{Introduction}

Can a variational quantum optimizer trained on yesterday’s problems produce high-quality parameters for tomorrow’s, possibly unexpected, optimization task? In realistic deployments, the eventual problem class may not be known when the optimizer is trained, or a new formulation may arise under time constraints that make full quantum–classical re-optimization impractical. This creates a pressing need for variational parameter-generation methods that can transfer across problem classes and adapt to new instances in only a few steps. 
Let us focus on the quantum approximate optimization algorithm (QAOA)~\cite{farhi2014qaoa,hadfield2019quantum,herrman2022multi} that is one of the leading approaches for combinatorial optimization problems, enabling a wide range of applications, such as finance~\cite{herman2023quantum}, biology~\cite{outeiral2021prospects}, and scientific computing~\cite{lin2022lecture}.

QAOA is a variational hybrid quantum–classical algorithm with a direct application to solving combinatorial optimization problems. 
It employs a parameterized quantum circuit whose angles are updated iteratively by a classical optimizer.
While the framework has advanced rapidly, efficiently tuning such parameters remains challenging.
As the circuit depth or the number of qubits grows, the optimization landscape can become exponentially flat, a phenomenon referred to as barren plateaus~\cite{mcclean2018barren, anschuetz2022quantum, wang2021noise, kulshrestha2204beinit}. 
Moreover, scaling QAOA to larger instances typically requires deeper circuits, which further complicates optimization and amplifies the impact of hardware noise on near-term devices~\cite{zhou2020quantum}.

\begin{figure}
    \centering
    \includegraphics[width=\linewidth]{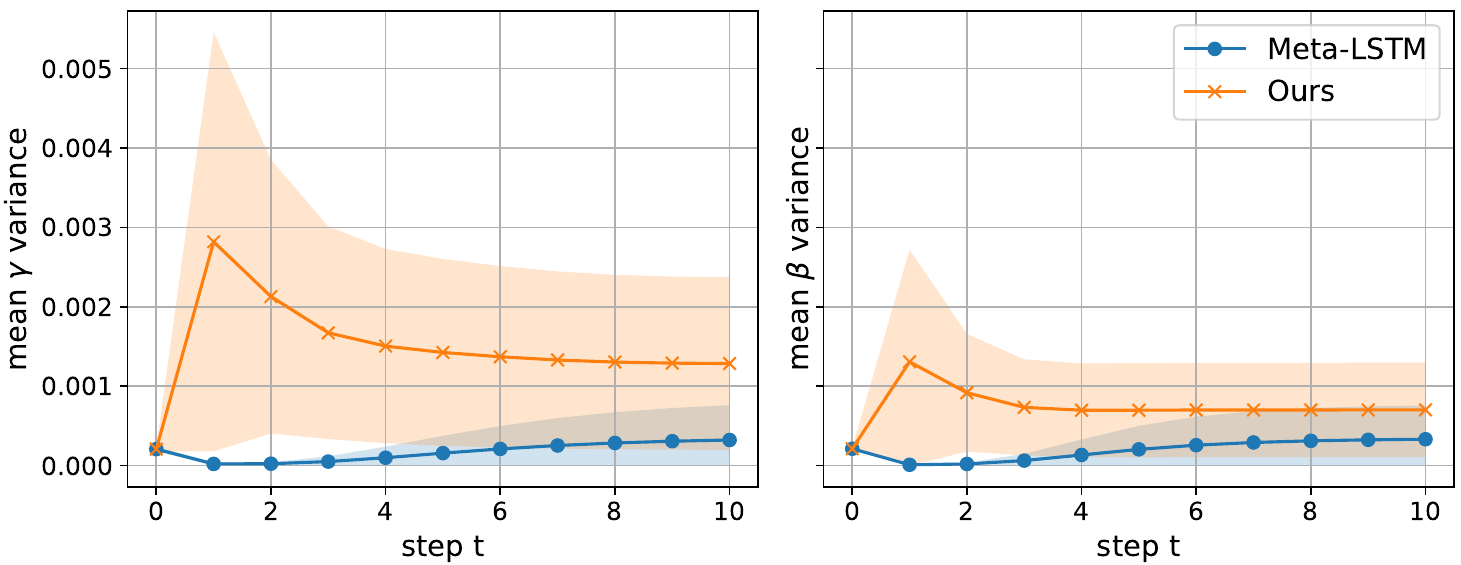}
    \caption{Diversity of generated parameter trajectories on QAOA circuits for MaxCut with depth $p=10$. At each rollout step $t$, we compute the mean squared deviation of the generated angles from the instance-mean trajectory, averaged over angle dimensions $\{\gamma_\ell\}_{\ell=1}^p$ (left) and $\{\beta_\ell\}_{\ell=1}^p$ (right). Curves show the mean across test instances; shaded bands indicate $\pm 1$ standard deviation across instances. Meta-LSTM exhibits lower diversity (near-identical paths) than our graph-conditioned meta-optimizer.}
    \label{fig:motif}
\end{figure}

These challenges have motivated a range of parameter initialization strategies, including linear-ramp schedules~\cite{zhou2020quantum,montanez2024towards}, multistart optimization~\cite{shaydulin2019multistart}, GPT-based \cite{tyagin2025qaoa}, and parameter transfer~\cite{montanez2402transfer,brandao2018fixed,galda2021transferability,falla2024graph,nguyen2025cross}.
A complementary approach is \emph{parameter generation}, in which a neural meta-optimizer is trained to propose good initial parameters~\cite{verdon2019learning,wilson2021optimizing,wang2021quantum,huang2022learning,friedrich2025learning}.
Verdon \emph{et al.}~\cite{verdon2019learning} proposed a lightweight Long Short-Term Memory (LSTM) network~\cite{hochreiter1997long} as a meta-optimizer that produces a parameter trajectory over a fixed horizon, mimicking the iterative updates of a classical optimization routine.
Huang \emph{et al.}~\cite{huang2022learning} subsequently refined the training objective and extended the framework to the variational quantum eigensolver for quantum chemistry applications.
However, our empirical results indicate that the meta-optimizer often produces \textit{near-identical parameter trajectories}, effectively collapsing to an average ``best” update path across the training instances.
This lack of expressivity limits the optimizer’s ability to adapt and consequently hinders generalization to more complex instances.
In this paper, we argue that incorporating instance-dependent conditioning is necessary to steer the optimizer toward more expressive trajectories and improve robustness across diverse instances (Figure~\ref{fig:motif}). 
Specifically, we demonstrate that conditioning the meta-optimizer on graph embeddings improves generalization to larger problems.

Furthermore, in many settings, closely related combinatorial formulations differ only in the objective or the addition of constraints. 
Re-optimizing QAOA parameters for each formulation can be expensive, motivating approaches that reuse learned optimization behavior across problem classes. 
Cross-problem transfer provides a mechanism to amortize parameter search by leveraging shared structure between objectives~\cite{montanez2402transfer,nguyen2025cross}.
While parameter transferability has been studied primarily within the same problem class, cross-problem transfer in QAOA remains relatively underexplored.
Nguyen \emph{et al.}~\cite{nguyen2025cross} show that cross-problem transfer based solely on structural similarity, e.g. using nearest neighbors on Graph2Vec features~\cite{narayanan2017graph2vec}, can be suboptimal.
To this end, a natural question arises: \textit{how can we learn graph embeddings whose features transfer across problem classes?} 
This motivates incorporating problem-specific information directly into the representation by encoding the objective and constraints, producing richer and more transferable features. 

\begin{figure}
    \centering
    \includegraphics[width=\linewidth]{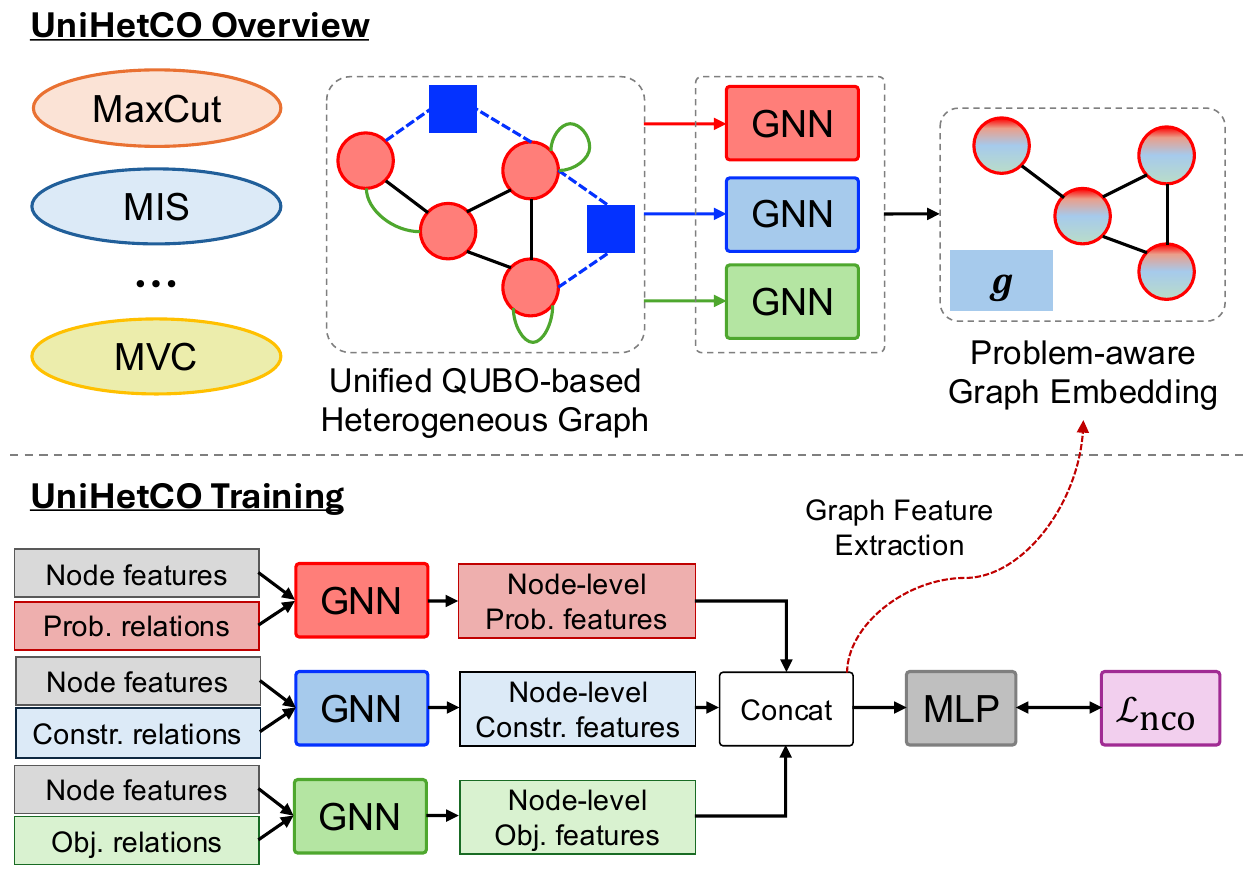}
    \caption{Overview of the UniHetCO graph encoder. UniHetCO not only encodes the problem instance structure (\textbf{\textcolor{red}{red}}), but also its objective (\textbf{\textcolor{ForestGreen}{green}}) and constraints (\textbf{\textcolor{blue}{blue}}), into a unified graph representation. This property makes feature transfer across different problem classes possible.}
    \label{fig:unihetco}
\end{figure}

Accordingly, we adopt the UniHetCO framework by Nguyen and Safro~\cite{nguyen2026unihetco}, which maps multiple problem classes into a unified quadratic programming (QP) formulation.
UniHetCO augments the original graph by adding objective-coupling relations, as well as introducing constraint nodes and variable–constraint edges (Figure~\ref{fig:unihetco}).
In other words, UniHetCO not only produces structure-aware but also \textit{problem-aware} embeddings, which accelerate cross-problem transfer.
It then learns graph embeddings by training a solver to minimize a loss that combines the objective value with constraint-violation penalties.
In our pipeline, we first pre-train the graph embeddings and then condition them on the training of the meta-optimizer for QAOA (Figure~\ref{fig:title}).

Finally, we comprehensively evaluate our graph-conditioned meta-optimizer on 16 single-problem and 48 cross-problem transfer settings. 
Experiments span four combinatorial optimization problems, Maximum Cut (MaxCut), Maximum Independent Set (MIS), Maximum Clique (MaxClique), and Minimum Vertex Cover (MVC), and four QAOA circuit depths.
Moreover, we demonstrate the expressivity of the approach by visualizing the variance of generated parameter trajectories over 100 random graph instances.

In summary, our contributions are threefold:
\begin{enumerate}
    \item We condition a learned QAOA meta-optimizer on graph embeddings with structure-aware (Graph2Vec) and problem-aware encodings (UniHetCO).
    \item We show that graph conditioning improves generalization to larger instances and enables few-shot transfer across problem classes.
    \item We quantify the increased expressivity of the conditioned meta-optimizer by visualizing the variance of generated parameter trajectories over many test instances and demonstrate the effectiveness of the proposed approach.
\end{enumerate}



\section{Related Work}
\subsection{Learning to Learn Variational Quantum Algorithms}
Learning to learn (L2L), or meta-learning, has been a long-standing research field in the machine learning community~\cite{andrychowicz2016learning,munkhdalai2017meta,finn2017model,raghu2019rapid,nguyen2024adaptive} and has recently been applied to quantum neural networks and quantum variational algorithms~\cite{verdon2019learning,wilson2021optimizing,wang2021quantum,huang2022learning,friedrich2025learning}.
In the context of variational quantum algorithms (VQAs), L2L methods typically train a classical model to propose circuit parameters that yield low energy with only a few refinement steps.
Rather than relying on hand-designed update rules, the meta-optimizer is trained end-to-end using feedback from the quantum objective, enabling amortized optimization and improved sample efficiency. 
This general framework has been explored in several settings, including learned parameter initialization, learned update rules, and the transfer of optimization behavior across related instances.

\subsection{Cross-Problem Parameter Transferability}
This emerging research direction studies how QAOA parameters optimized for one problem class can be reused to accelerate optimization on another.
A common strategy is to leverage parameters learned on a more QAOA-native objective (e.g., MaxCut) and transfer them to a more challenging objective (e.g., constrained MIS), with the goal of reducing optimization effort while maintaining solution quality.
Monta\~nez \emph{et al.}~\cite{montanez2402transfer} present a numerical study of transferring QAOA parameters between different combinatorial optimization problems. 
While their results suggest the promise of cross-problem transfer, their empirical analysis is limited to a small set of randomly generated instances.
To this end, Nguyen \emph{et al.}~\cite{nguyen2025cross} have adopted a machine learning-based approach to systematically identify good donor candidates on MaxCut and transfer their parameters to the corresponding acceptors on MIS.
\section{Background}

\subsection{Meta-Optimizer for QAOA Parameter Generation}
The ``learning to learn'' (L2L) framework involves training a neural network to act as a meta-optimizer for QAOA. 
Concretely, an LSTM~\cite{hochreiter1997long} is employed to generate a short sequence of QAOA parameters that serve as strong initializations after only a few steps, following prior work on learned QAOA optimizers~\cite{verdon2019learning,huang2022learning}. 
At each rollout step, the LSTM maintains an internal memory and receives feedback about the current solution quality. 
The input at step $t$ consists of (i) the energy value achieved by the previous QAOA parameters and (ii) the previous parameters themselves. 
Using this information, the LSTM proposes an updated set of QAOA angles for the next step. 
Repeating this procedure for a fixed horizon produces a parameter trajectory that mimics an optimization process, but is generated by the learned model rather than by a hand-crafted optimizer.
The set of parameters generated at the last unrolling step is then used to evaluate the QAOA circuit expectation value.

\subsection{Graph Embedding}
Graph embedding refers to a broad class of techniques that map vertices, edges, or entire graphs to low-dimensional vector representations.
These embeddings capture key structural and semantic information from the graphs, making them suitable for downstream machine learning tasks.
Existing approaches can be divided into node-level embedding~\cite{sybrandt2019fobe,ding2020unsupervised,grover2016node2vec} and graph-level embedding~\cite{wang2021graph,cai2018simple,galland2019invariant,narayanan2017graph2vec}.
Node-level embedding methods assign a vector to each vertex while preserving local structure, such as neighborhood relationships and attribute similarity, making them well suited for node-centric tasks like classification, link prediction, and community detection.
Graph-level embedding techniques produce a single fixed-size vector for an entire graph that summarizes global organization and semantics, which is more appropriate for graph-level objectives such as classification and regression.

\section{Graph-Conditioned Meta-Optimizer}

\begin{figure*}[ht]
    \centering
    \includegraphics[width=0.85\linewidth]{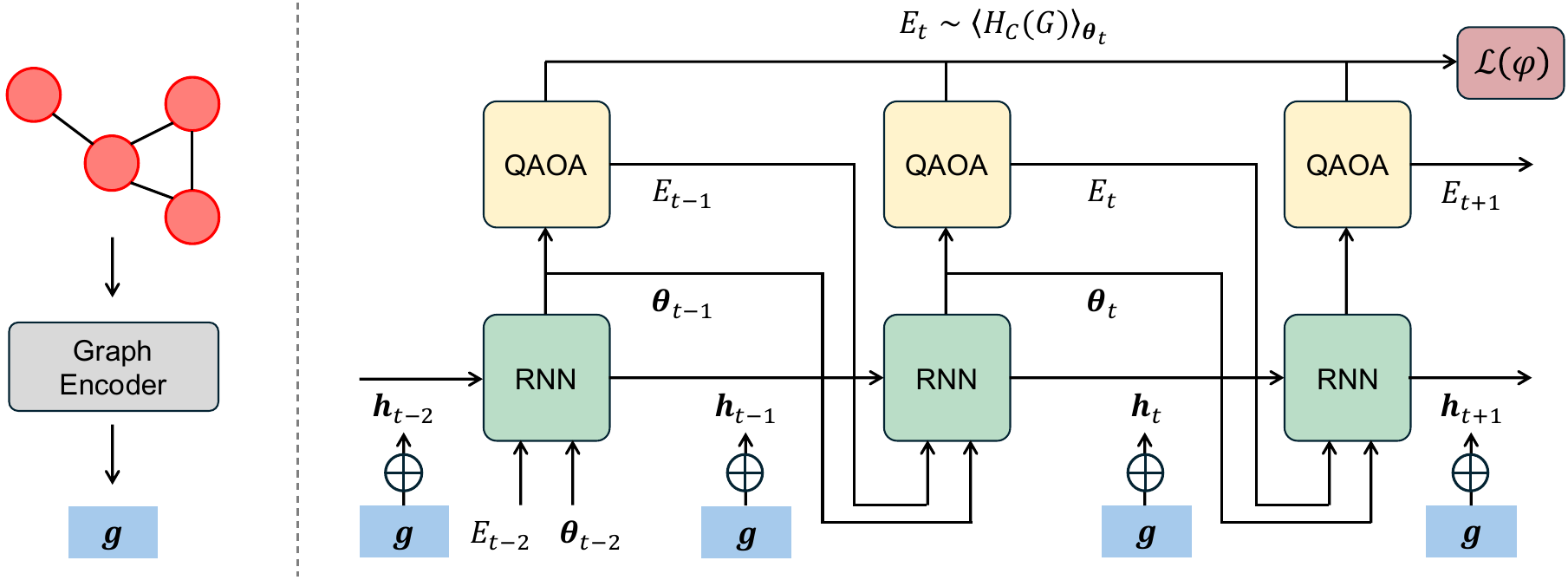}
    \caption{Overview of the graph-conditioned meta-optimizer training pipeline. \textbf{(Left)} The graph instance is first projected to a fixed-dimension vector $\mathbf{g}$ by a graph encoder. \textbf{(Right)} A meta-optimizer, realized by a recurrent neural network (RNN), is trained to produce a trajectory of QAOA variational parameters. At each step $t$, the graph embedding $\mathbf{g}$ is injected to the RNN hidden state $\mathbf{h}_t$, effectively conditioning parameter generation on instance-specific graph features.}
    \label{fig:title}
\end{figure*}

We extend the L2L framework by training a neural meta-optimizer that produces QAOA parameters conditioned on the problem instance.
Given a graph instance, we compute a $d$-dimensional embedding $\mathbf{g} = \Phi(G)\in\mathbb{R}^d$, where $\Phi$ is the graph encoder, and use it to condition an optimizer network $f_{\boldsymbol{\varphi}}$, parameterized by the weights $\boldsymbol{\varphi}$.
The optimizer generates a short parameter trajectory $\{\boldsymbol{\theta}_t\}_{t=1}^T$, where $\boldsymbol{\theta}_t=(\boldsymbol{\gamma}_t,\boldsymbol{\beta}_t)\in\mathbb{R}^{2p}$ denotes the QAOA angles, with a depth $p$ (Figure~\ref{fig:title}).

In particular, we model trajectory generation as a recurrent optimization procedure.
Let $E(\boldsymbol{\theta};G)=\bra{\psi(\boldsymbol{\theta})}H_C(G)\ket{\psi(\boldsymbol{\theta})}$ be the QAOA energy of graph $G$. 
At step $t$, the meta-optimizer takes in the QAOA parameters and energy from the previous step $t-1$, and outputs the next parameters:
\begin{align}
    (\mathbf{h}_t, \mathbf{s}_t) = f_{\boldsymbol{\varphi}}(\mathbf{z}_t,\mathbf{h}_{t-1},\mathbf{s}_{t-1}),\quad \boldsymbol{\theta}_t=\mathbf{W}\mathbf{h}_t,
\end{align}
where $\mathbf{z}_t = [E(\boldsymbol\theta_{t-1};G), \boldsymbol\theta_{t-1}]$ denotes the input at step $t$, $[\cdot,\cdot]$ denotes concatenation, $\mathbf{h}_t$ and $\mathbf{s}_t$ are the hidden and cell states, respectively.
The recurrent module is implemented as an LSTM network~\cite{hochreiter1997long}, which is designed to mitigate vanishing gradients and improve stability when training recurrent models via backpropagation through time.
We briefly summarize the LSTM update at step $t$. Given input $\mathbf{z}_t$, previous hidden state $\mathbf{h}_{t-1}$, and cell state $\mathbf{s}_{t-1}$, the LSTM computes
\begin{align}
\mathbf{i}_t &= \sigma(\mathbf{W}_{ii} \mathbf{z}_t + \mathbf{W}_{hi} \mathbf{h}_{t-1}), \\
\mathbf{f}_t &= \sigma(\mathbf{W}_{if} \mathbf{z}_t + \mathbf{W}_{hf} \mathbf{h}_{t-1}), \\
\tilde{\mathbf{s}}_t &= \tanh(\mathbf{W}_{is} \mathbf{z}_t + \mathbf{W}_{hs} \mathbf{h}_{t-1}), \\
\mathbf{o}_t &= \sigma(\mathbf{W}_{io} \mathbf{z}_t + \mathbf{W}_{ho} \mathbf{h}_{t-1}),
\end{align}
where $\sigma(\cdot)$ denotes the sigmoid function. We omit the bias terms for simplicity of notations. The cell and hidden states are then updated as
\begin{align}
\mathbf{s}_t &= \mathbf{f}_t \odot \mathbf{s}_{t-1} + \mathbf{i}_t \odot \tilde{\mathbf{s}}_t, \\
\mathbf{h}_t &= \mathbf{o}_t \odot \tanh(\mathbf{s}_t),
\end{align}
with $\odot$ denoting element-wise multiplication (or Hadamard product). The gating vectors $(\mathbf{i}_t,\mathbf{f}_t,\mathbf{o}_t)$ regulate information flow through the cell state, enabling more stable credit assignment over long horizons.

The optimizer $f_{\boldsymbol\varphi}$ learns to minimize a loss function $\mathcal{L}(f_{\boldsymbol\varphi})$ derived from the energy trajectory $E_1\dots,E_T$.
We adopt the loss function from Huang \emph{et al.}~\cite{huang2022learning} which is a decay weighted sum over the trajectory of the normalized energies:
\begin{align}
    \label{eq:loss-function}
    \mathcal{L}(f_\varphi) = \sum_t \omega_t\bar E_t,
\end{align}
where $\{\omega_t\}_{t=1}^{T}$ assigns larger weights to later rollout steps to emphasize the quality of the final parameter proposals. As different instances may induce different magnitudes of the QAOA energy, we normalize it by the $\ell_1$ norm of the Pauli coefficients to stabilize the gradient updates.
Given a cost Hamiltonian with Pauli decomposition
\begin{align}
    H_C = \sum_{j} \boldsymbol\alpha_j P_j,
\end{align}
we define the Pauli coefficient norm as
\begin{align}
    \|H_C\|_1 \triangleq \|\boldsymbol{\alpha}\|_1 = \sum_{j} |\boldsymbol\alpha_j|.
\end{align}
We then use the normalized energy
\begin{align}
    \bar{E}(\boldsymbol\theta;G) \triangleq \frac{E(\boldsymbol\theta;G)}{\|H_C(G)\|_1},
\end{align}
which is bounded in $[-1,1]$ since each Pauli string $P_j$ has eigenvalues $\pm 1$.
We update the meta-optimizer parameters $\boldsymbol{\varphi}$ using gradient descent:
\begin{align}
    \boldsymbol{\varphi}' \leftarrow \boldsymbol{\varphi} - \eta \nabla_{\boldsymbol{\varphi}}\mathcal{L},
\end{align}
where $\eta$ is the learning rate. To compute $\nabla_{\boldsymbol{\varphi}}\mathcal{L}$, we differentiate through the unrolled meta-optimization trajectory. At each rollout step $t$, the loss depends on the QAOA energy
\begin{align}
    E_t = \langle H_C \rangle_{\boldsymbol\theta_t} 
    = \bra{\psi(\boldsymbol\theta_t)} H_C \ket{\psi(\boldsymbol\theta_t)}.
\end{align}
The gradient of the energy with respect to a circuit parameter $\boldsymbol\theta_{t,k}$ is computed using adjoint differentiation method by Jones and Gacon~\cite{jones2020efficient}. Concretely, given the QAOA energy $E_t(\boldsymbol\theta_t)=\langle H_C\rangle_{\boldsymbol\theta_t}$, the simulator evaluates the gradient $\nabla_{\boldsymbol\theta_t}E_t$ via an adjoint-backpropagation procedure, which backpropagates through the circuit using an adjoint state and returns $\partial E_t/\partial \boldsymbol\theta_{t,k}$ without requiring parameter shifts~\cite{crooks2019gradients, schuld2019evaluating}.
Gradients are then propagated to the meta-optimizer parameters via the chain rule:
\begin{align}
    \nabla_{\boldsymbol{\varphi}} \mathcal{L} = \sum_{t=1}^{T} \frac{\partial \mathcal{L}}{\partial E_t} \cdot \frac{\partial E_t}{\partial \boldsymbol\theta_t} \cdot \frac{\partial \boldsymbol\theta_t}{\partial \boldsymbol\varphi},
\end{align}
where $\frac{\partial \boldsymbol\theta_t}{\partial \boldsymbol\varphi}$ is obtained by backpropagating through the unrolled LSTM that generates $\{\boldsymbol\theta_t\}_{t=1}^T$.
In mini-batch training, we sample a batch $\mathcal{B}$ of instances and replace $\mathcal{L}$ with the average batch loss,
\begin{align}
    \mathcal{L}_{\mathcal{B}} =\frac{1}{|\mathcal{B}|}\sum_{G\in \mathcal{B}} \mathcal{L}(G),
\end{align}
so that each update uses $\nabla_{\boldsymbol{\varphi}}\mathcal{L}_{\mathcal{B}}$.

\vspace{5pt}
\noindent\textbf{Graph conditioning at each step.} To ensure that the instance information remains available throughout the unrolled horizon, we inject the graph embedding at every step by augmenting the recurrent hidden states $h_t$,
\begin{align}
    \tilde{\mathbf{h}}_t = \mathbf{h}_t + \mathbf{g}, \quad \boldsymbol\theta_t = \mathbf{W}\tilde{\mathbf{h}}_t.
\end{align}

\section{Problem-aware Graph Embedding}
Building a graph embedding framework with strong feature transfer is the key to enabling \textit{generalization across larger problem sizes and different problem classes}.
To this end, we employ UniHetCO to construct a problem-aware graph representation grounded in the general Quadratic Programming (QP) formulation and learn transferable embeddings via unsupervised neural combinatorial optimization (NCO)~\cite{nguyen2026unihetco}.
Unsupervised NCO trains a neural solver without labeled optimal solutions by directly minimizing an instance objective on the model’s predictions.
We refer the reader to Nguyen and Safro~\cite{nguyen2026unihetco} for further details on the UniHetCO framework.

Unlike existing graph embedding methods, such as Graph2Vec~\cite{narayanan2017graph2vec}, which focus solely on the structure of the graph instance, UniHetCO builds a representation that casts various node subset-selection problems into a unified learning objective.
Given an instance with $N$ decision variables, we first write it in a common QP template with quadratic and linear objective terms and linear constraints:
\begin{align}
    \min_{\mathbf{x}\in \mathbb{R}^n} \frac{1}{2}\mathbf{x}^\top \mathbf{Q}\mathbf{x} + \mathbf{c}^\top\mathbf{x}
    \quad \mathrm{s.t. } \mathbf{A}\mathbf{x} \leq \mathbf{b},
\end{align}
where $\mathbf{x}\in \mathbb{R}^n$ is a vector of decision variables, $\mathbf{Q}\in\mathbb{R}^{n\times n}$ is a coefficient matrix, $\mathbf{c}\in\mathbb{R}^n$ denotes linear coefficients, and $\mathbf{A}\mathbf{x}\le \mathbf{b}$ are optional linear inequality constraints with $\mathbf{A}\in\mathbb{R}^{m\times n}$ and $\mathbf{b}\in\mathbb{R}^m$.
By absorbing $\mathbf{c}$ into the diagonal of $\mathbf{Q}$ and converting the linear constraints into a penalty term, we obtain an equivalent Quadratic Unconstrained Binary Optimization (QUBO) problem of the form:
\begin{align}
    \min_{\mathbf{x}\in\{0,1\}^n} \mathbf{x}^\top \tilde{\mathbf{Q}}\,\mathbf{x} + \lambda\cdot \text{Penalty}(\mathbf{x}), \quad \tilde{\mathbf{Q}}=\mathbf{Q}+\mathrm{diag}(\mathbf{c}).
\end{align}

UniHetCO then encodes the resulting objective and constraints \emph{directly into the input graph} so that a single neural model can be trained across problem classes with one unified, label-free objective.

\subsection{Unified Heterogeneous Graph Representation}
For each instance, UniHetCO constructs a heterogeneous graph with (i) decision variable nodes $\mathcal{V}_\mathrm{var}$, (ii) constraint nodes $\mathcal{V}_\mathrm{constr}$, and three relation types:
\paragraph{Problem graph} $G_\mathrm{prob}=(\mathcal{V}_\mathrm{var},\mathcal{E}_\mathrm{prob})$ contains edges capturing the original instance structure.
\paragraph{Objective graph} $G_\mathrm{obj}=(\mathcal{V}_\mathrm{var},\mathcal{E}_\mathrm{obj},w_\mathrm{obj})$ captures the objective-coupling relations, where $\mathcal{E}_\mathrm{obj}=\mathcal{E}_\mathrm{off} \cup \mathcal{E}_\mathrm{diag}$. 
Off-diagonal edges represent the quadratic terms $\mathcal{E}_{\mathrm{off}}=\{(u, v): u < v, \mathbf{Q}_{uv} \neq 0\}, w_{uv} = \mathbf{Q}_{uv}$, and self-loops representing the linear terms in $\mathbf{c}$ that are absorbed into the diagonal of $\mathbf{Q}$, $\mathcal{E}_{\mathrm{diag}}=\{(u, u): u \in \mathcal{V}_\mathrm{var}, \mathbf{Q}_{uu} + \mathbf{c}_u \neq 0\}, w_{uu} = \mathbf{Q}_{uu} + \mathbf{c}_{uu}$.
\paragraph{Constraint hypergraph} encodes the variable-constraint relations, where the variables appearing in a constraint form a hyperedge.
UniHetCO first standardizes all linear inequality constraints to the ``less-than-or-equal'' form by multiplying any inequality of the form $\alpha^\top x \geq \beta$ by $-1$.
It then constructs the constraint graph via a bipartite star expansion, connecting $\mathcal{V}_\mathrm{var}$ to $\mathcal{V}_\mathrm{constr}$, with each constraint node storing its right-hand-side values. This yields $G_\mathrm{constr}=(\mathcal{V}_\mathrm{var},\mathcal{V}_\mathrm{constr},\mathcal{E}_\mathrm{constr})$.

This representation unifies MaxCut, MIS, MaxClique, MVC, and MDS within the same heterogeneous graph input format $G = (\mathcal{V}_\mathrm{var},\mathcal{V}_\mathrm{constr},\mathcal{E}_\mathrm{prob},\mathcal{E}_\mathrm{obj},\mathcal{E}_\mathrm{constr})$.

\subsection{Heterogeneous Graph Neural Network Solver}
Given the heterogeneous input, UniHetCO applies relation-specific message passing to compute node-level representations for each relation type. 
Concretely, for each variable node $v\in\mathcal{V}_{\mathrm{var}}$, a separate message-passing network aggregates information from its neighbors under the problem, objective, and constraint relations, yielding three embeddings
$\mathbf{h}^{\mathrm{prob}}_v$, $\mathbf{h}^{\mathrm{obj}}_v$, and $\mathbf{h}^{\mathrm{constr}}_v$.
Each embedding captures complementary signals: $\mathbf{h}^{\mathrm{prob}}_v$ summarizes the original instance structure, $\mathbf{h}^{\mathrm{obj}}_v$ encodes quadratic and linear objective couplings, and $\mathbf{h}^{\text{constr}}_v$ propagates constraint-specific context through variable--constraint interactions.
The three embeddings are then fused, by concatenation followed by a multi-layer perceptron, to form a unified representation $\mathbf{h}_v$, and a shared prediction head maps $\{\mathbf{h}_v\}_{v=1}^n$ to a relaxed selection vector $\mathbf{x}\in[0,1]^n$, where each entry $\mathbf{x}_v$ represents the predicted likelihood that node $v$ is included in the solution.
At inference time, a greedy decoder projects the relaxed output back to a feasible discrete solution.

\subsection{Universal QUBO Training Objective}
Training uses a single surrogate objective shared across problem classes, consisting of an objective term derived from the encoded quadratic form and a penalty on constraint violations. 
The loss function for the objective graph $(\mathcal{V}_\mathrm{var},\mathcal{E}_\mathrm{obj})$ is as follows:
\begin{align}
    \mathcal{L}_{\mathrm{obj}}(G) = \mathbf{x}^\top \tilde{\mathbf{Q}}\mathbf{x},
\end{align}
From the constraint graph $(\mathcal{V}_\mathrm{var},\mathcal{V}_\mathrm{constr},\mathcal{E}_\mathrm{constr})$, we define a hinge loss over the violation vector $\mathbf{A}\mathbf{x}-\mathbf{b}$:
\begin{align}
    \mathcal{L}_{\mathrm{constr}}(G)
= \mathbf{1}^\top \max\big(\mathbf{0}, \mathbf{A}\mathbf{x} - \mathbf{b}\big),
\end{align}
where $\mathbf{1}\in\mathbb{R}^m$ is the all-one vector.
The model learns to minimize the objective loss while reducing violations of linear constraints, without requiring ground-truth solutions:
\begin{align}
    \mathcal{L}_\mathrm{NCO}(G) = \lambda_\mathrm{obj}\mathcal{L}_\mathrm{obj}(G) + \lambda_\mathrm{constr}\mathcal{L}_\mathrm{constr}(G),
\end{align}
where $\lambda_\mathrm{obj}$ and $\lambda_\mathrm{constr}$ are the balancing coefficients. Empirically, we set $\lambda_\mathrm{obj}=\lambda_\mathrm{constr}=1.0$.

\subsection{Graph Embedding Extraction}
After training, we first extract the heterogeneous embeddings of the individual nodes $\mathbf{h}_v=[\mathbf{h}^{\mathrm{prob}}_v,\mathbf{h}^{\mathrm{obj}}_v,\mathbf{h}^{\mathrm{constr}}_v]$ from the heterogeneous GNN $\Phi$.
We then obtain a graph-level representation by mean pooling over the node embeddings:
\begin{align}
    \mathbf{g} = \frac{1}{|\mathcal{V}_\mathrm{var}|} \sum_v \mathbf{h}_v.
\end{align}
\section{Empirical Results}
\subsection{Experiment Setup}
\paragraph{Evaluation Metrics} We evaluate the baselines on several metrics: (i) feasibility rate, (ii) optimal hit rate, and (iii) approximation ratio.

\noindent\textbf{Feasibility Rate.}
After obtaining the final quantum state $\ket{\psi(\boldsymbol{\theta}^*)}$, we perform 5000 measurement shots and collect the resulting bit strings, denoted as the set $\mathcal{S}$.
For constrained problem classes, we compute the feasibility rate as the fraction of sampled bit strings that satisfy the problem constraints.

\begin{align}
\mathrm{FR} = \frac{1}{|\mathcal{S}|}\sum_{\mathbf{x}\in \mathcal{S}}\mathbb{I}[\mathbf{x} \in \mathcal{F}],
\end{align}
where $\mathcal{F}\subseteq \mathcal{S}$ denotes the feasible set and $\mathbb{I}[\cdot]$ is the indicator function.

\noindent\textbf{Approximation Ratio.}
The approximation ratio is defined as the ratio of the expectation value of the QAOA circuit to the optimal solution value.
Given the optimized variational parameters $\boldsymbol\theta^*$, the expectation value is calculated as
\begin{align}
    \mathrm{AR} = \braket{H_C(G;\boldsymbol\theta^*)} &\coloneq \bra{\psi(\boldsymbol\theta^*)} H_C(G) \ket{\psi(\boldsymbol\theta^*)}\label{eq:expval}\\ 
    &= \sum_{\mathbf{x}\in \{0,1\}^n} \mathcal{C}(\mathbf{x}) \text{Pr}(\mathbf{x}),\label{eq:emp-expval}
\end{align}
where $\text{Pr}(\mathbf{x})$ is the probability of observing $\mathbf{x}$ when measuring $\ket{\psi(\boldsymbol\theta^*)}$.
For unconstrained problem classes, we report the expectation value in~\eqref{eq:expval}.
For constrained problem classes, we estimate an empirical expectation value conditioned on feasibility by restricting to sampled bitstrings in the feasible set $\mathcal{F}$, extending Eq.~\eqref{eq:emp-expval} as:
\begin{align}
    \braket{H_C(G;\boldsymbol\theta^*)} \coloneq \sum_{\mathbf{x}\in \mathcal{F}} \mathcal{C}(\mathbf{x})\,\Pr(\mathbf{x}).
    \label{eq:emp-expval-feasible}
\end{align}
Note that for minimization problems, e.g. MVC, we report the $\mathrm{AR}-1$ metric, effectively the relative gap above the optimum, with lower value corresponding to better performance.

\begin{table*}[ht]
    \centering
    \caption{Results of single-problem setting on the MaxCut, MIS and MaxClique problem classes. Best results are highlighted in bold. Asterisk (*) denotes better results than Vanilla QAOA.}
    \resizebox{\linewidth}{!}{
    \begin{tabular}{l|cccc|cccc|cccc|cccc}
         \toprule
         Circuit depth & \multicolumn{4}{c|}{$p=4$} & \multicolumn{4}{c|}{$p=6$} & \multicolumn{4}{c|}{$p=8$} & \multicolumn{4}{c}{$p=10$} \\
         \midrule
         Metrics & $p(\mathbf{x}^*)$ & AR & FR & $T$ & $p(\mathbf{x}^*)$ & AR & FR & $T$ & $p(\mathbf{x}^*)$ & AR & FR & $T$ & $p(\mathbf{x}^*)$ & AR & FR & $T$\\
         \midrule
         \multicolumn{17}{c}{\textbf{Maximum Cut}}\\
         \midrule
         Vanilla QAOA & 25.09 & 93.80 & - & 231.44 & 40.33 & 96.01 & - & 296.74 & 53.38 & 97.22 & - & 357.60 & 66.14 & 98.16 & - & 405.29 \\
         Meta-LSTM & 18.99 & 92.02 & - & 10.00 & 37.24 & 95.01 & - & 10.00 & \textbf{42.71} & 95.12 & - & 10.00 & 55.59 & 96.68 & - & 10.00 \\
         \midrule
         G2V-Meta-LSTM & 17.47 & 91.80 & - & 10.00 & 27.33 & 93.76 & - & 10.00 & 30.52 & 94.40 & - & 10.00 & 45.69 & 95.77 & - & 10.00 \\
         Uni-Meta-LSTM & \textbf{20.95} & \textbf{92.63} & - & 10.00 & \textbf{37.27}& \textbf{95.05} & - & 10.00 & 41.23 & \textbf{95.76} & - & 10.00 & \textbf{60.84} & \textbf{97.16} & - & 10.00 \\
         \midrule
         \multicolumn{17}{c}{\textbf{Maximum Independent Set}}\\
         \midrule
         Vanilla QAOA & 20.32 & 71.05 & 79.62 & 401.03 & 33.65 & 80.34 & 76.88 & 434.78 & 44.52 & 85.07 & 80.63 & 467.31 & 54.22 & 87.94 & 83.39 & 477.91 \\
         Meta-LSTM & 14.77 & 63.59 & 79.03 & 10.00 & 32.57 & \textbf{82.43}* & 66.60 & 10.00 & 39.72 & 88.13* & 65.17 & 10.00 & 41.37 & 84.00 & 74.09 & 10.00 \\
         \midrule
         G2V-Meta-LSTM & 17.78 & \textbf{70.93} & 70.11 & 10.00 & 20.21 & 73.51 & 66.79 & 10.00 & 44.66 & 82.77 & 88.60 & 10.00 & 42.10 & \textbf{86.63} & 71.37 & 10.00 \\
         Uni-Meta-LSTM & \textbf{18.49} & 70.49 & 80.07 & 10.00 & \textbf{33.79}* & 81.09* & 78.30 & 10.00 & \textbf{50.17}* & \textbf{89.05}* & 79.25 & 10.00 & \textbf{46.08} & 84.31 & 89.66 & 10.00 \\
         \midrule
         \multicolumn{17}{c}{\textbf{Maximum Clique}}\\
         \midrule
         Vanilla QAOA & 27.74 & 80.34 & 75.79 & 338.72 & 41.73 & 87.05 & 76.76 & 401.56 & 54.81 & 90.61 & 82.77 & 438.06 & 64.53 & 92.83 & 87.06 & 460.58 \\
         Meta-LSTM & 11.57 & 65.45 & 53.66 & 10.00 & 33.78 & 86.99 & 61.93 & 10.00 & 33.60 & 87.25 & 59.54 & 10.00 & 39.70 & 87.96 & 67.39 & 10.00 \\
         \midrule
         G2V-Meta-LSTM & 21.67 & 76.88 & 74.29 & 10.00 & 38.00 & 83.77 & 81.83 & 10.00 & 40.56 & 89.63 & 67.24 & 10.00 & 48.87 & 88.37 & 85.86 & 10.00 \\
         Uni-Meta-LSTM & \textbf{23.23} & \textbf{77.45} & 79.60 & 10.00 & \textbf{39.78} & \textbf{87.46}* & 73.95 & 10.00 & \textbf{54.82}* & \textbf{90.87}* & 83.30 & 10.00 & \textbf{50.74} & \textbf{88.39} & 80.66 & 10.00 \\
         \bottomrule
    \end{tabular}
    }
    \label{tab:maxcut}
\end{table*}

\begin{table*}[ht]
    \centering
    \caption{Results of single-problem setting on the MVC problem class. Best results are highlighted in bold. Asterisk (*) denotes better results than Vanilla QAOA.}
    \resizebox{\linewidth}{!}{
    \begin{tabular}{l|cccc|cccc|cccc|cccc}
         \toprule
         Circuit depth & \multicolumn{4}{c|}{$p=4$} & \multicolumn{4}{c|}{$p=6$} & \multicolumn{4}{c|}{$p=8$} & \multicolumn{4}{c}{$p=10$} \\
         \midrule
         Metrics & $p(\mathbf{x}^*)$ & $\mathrm{AR}-1$ & FR & $T$ & $p(\mathbf{x}^*)$ & $\mathrm{AR}-1$ & FR & $T$ & $p(\mathbf{x}^*)$ & $\mathrm{AR}-1$ & FR & $T$ & $p(\mathbf{x}^*)$ & $\mathrm{AR}-1$ & FR & $T$\\
         \midrule
         Vanilla QAOA & 20.37 & 12.55 & 80.55 & 403.86 & 33.44 & 8.23 & 77.70 & 440.75 & 45.79 & 6.20 & 82.02 & 463.46 & 54.40 & 5.00 & 83.14 & 468.42 \\
         Meta-LSTM & 14.80 & 15.59 & 79.16 & 10.00 & 32.69 & 8.18* & 66.70 & 10.00 & 39.80 & \textbf{5.64}* & 65.20 & 10.00 & 41.42 & 6.66 & 74.12 & 10.00 \\
         \midrule
         G2V-Meta-LSTM & 17.84 & 14.04 & 70.10 & 10.00 & 20.36 & 11.69 & 66.85 & 10.00 & \textbf{44.73} & 7.65 & 88.58 & 10.00 & 42.21 & 5.91 & 71.42 & 10.00 \\
         Uni-Meta-LSTM & \textbf{20.03} & \textbf{12.70} & 80.47 & 10.00 & \textbf{34.01}* & \textbf{8.14}* & 74.79 & 10.00 & 43.17 & 5.91* & 73.35 & 10.00 & \textbf{57.32}* & \textbf{4.56}* & 86.10 & 10.00 \\
         \bottomrule
    \end{tabular}
    }
    \label{tab:mvc}
\end{table*}

\noindent\textbf{Optimal Hit Rate.}
The optimal hit rate is defined as the ratio of sampled feasible bit strings that achieve the optimal value.
\begin{align}
    p(\mathbf{x}^*) = \frac{1}{\vert\mathcal{S}\rvert}\sum_{\mathbf{x}\in\mathcal{S}} \mathbb{I}[\mathcal{C}(\mathbf{x})=\mathcal{C}^* \wedge \mathbf{x}\in\mathcal{F}]
\end{align}
where $\mathcal{C}^*$ denotes the optimal objective value. 
This metric is particularly appropriate for constrained problems because it evaluates end-to-end solver performance under the same sampling budget for all methods. 
In contrast, reporting approximation ratios conditioned on feasibility can obscure differences in feasibility rates across methods. 
Therefore, by directly quantifying the chance of obtaining an optimal feasible solution within $|\mathcal{S}|$ shots, the optimal hit rate provides a fair and practically meaningful comparison when all approaches are evaluated on the same QAOA circuit family and the same measurement budget.

\paragraph{Dataset}
To create the graph dataset, we follow the experiment setting similar to that of~\cite{verdon2019learning,huang2022learning}, sampling 1000 random connected graphs for training and 100 for testing.
Graph sizes for training are drawn from $n\in[6,10]$, and edges are generated with probability $k/n$, where $k\in[3,n-1]$.
For testing, we fix the graph size to $n=12$ and use the same data generation procedure.
All graphs are non-isomorphic.

\paragraph{Baselines}
We compare our graph-conditioned meta-optimizer against three baselines. 
(i) \textbf{Vanilla QAOA}: for each instance, we optimize QAOA parameters from a random initialization $\boldsymbol{\theta}$ for up to $T=500$ steps; we treat this per-instance optimization as an empirical upper bound.
(ii) \textbf{Meta-LSTM}: an unconditioned LSTM meta-optimizer trained to generate QAOA parameter trajectories, following~\cite{verdon2019learning,huang2022learning}. 
(iii) \textbf{G2V-Meta-LSTM}: the same LSTM meta-optimizer conditioned on the Graph2Vec embedding.
We dub our method \textbf{Uni-Meta-LSTM}.

\paragraph{Implementation Details}
We set the rollout horizon for all meta-optimizer baselines to $T=10$ and $\omega_t=\frac{t}{10}$ for Eq.~\eqref{eq:loss-function}.
During training, we use a mini-batch of $\mathcal{B}=32$ samples per gradient update and train the meta optimizer for 100 epochs.
We set the LSTM hidden state and cell state dimension to 48.
We select the best checkpoint as the model that achieves the lowest mean energy on the training set and then deploy it on the test set.
For each problem class, we train a separate model at each circuit depth $p\in\{4,6,8,10\}$.
We implement quantum simulation in PennyLane~\cite{bergholm2022pennylane} and use PyTorch~\cite{paszke2019pytorch} to train the neural meta-optimizer.
We use the Adam optimizer~\cite{kingma2014adam} to train both the Vanilla QAOA and meta-learning-based methods, setting the learning rate to 0.01 for the former and 0.001 for the latter.
We conduct all experiments on the same seed using a single 16GB NVIDIA T4 GPU and run quantum simulations using PennyLane's \texttt{lightning.qubit} device.

\begin{figure*}[ht]
    \centering
    \includegraphics[width=\linewidth]{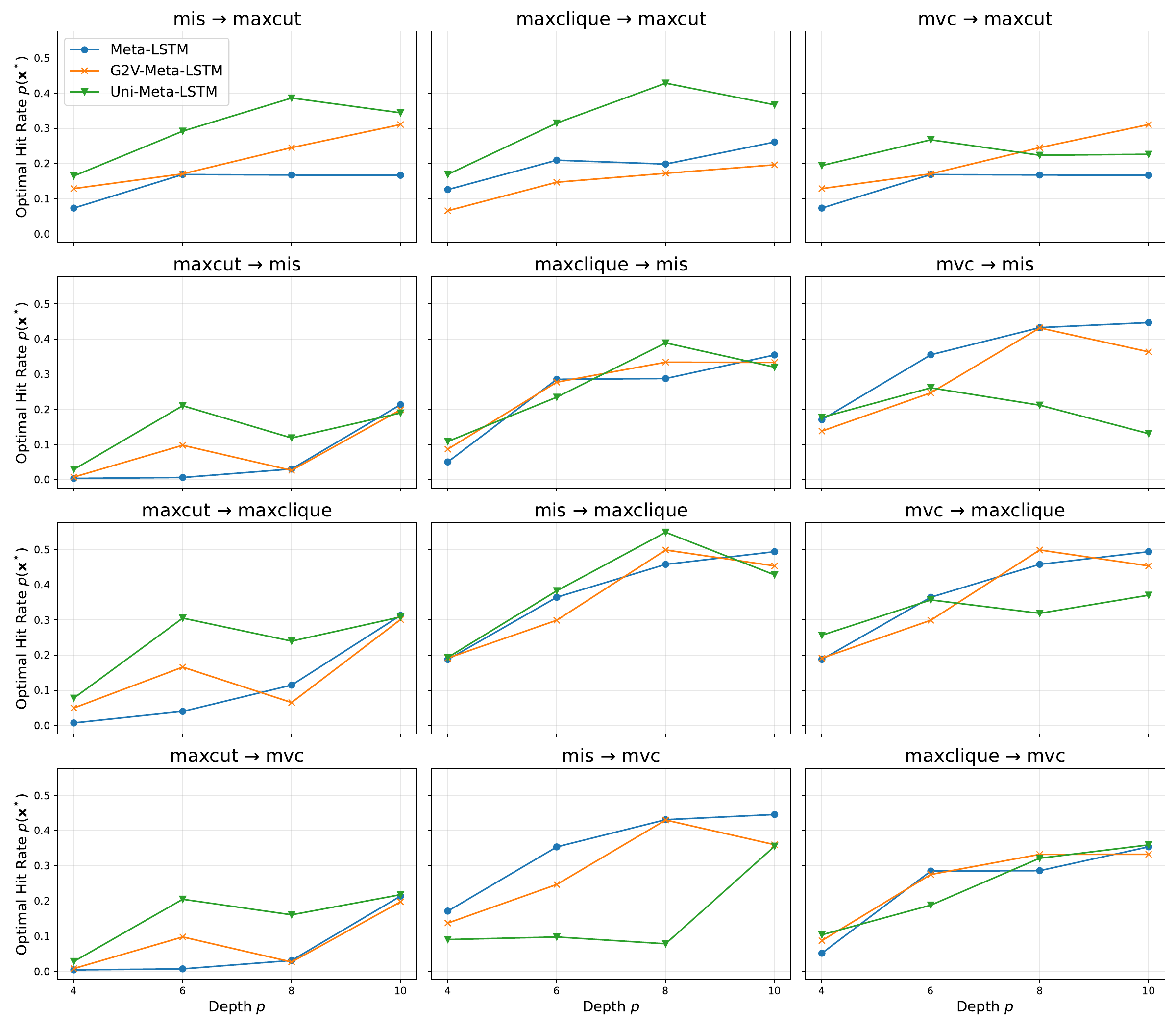}
    \caption{Optimal hit rate $p(\mathbf{x}^*)$ on cross-problem parameter transfer. We report pairwise problem transfer across four circuit depths $p\in\{4,6,8,10\}$, resulting in 48 transfer settings. The meta-optimizer is trained on one problem class and then fine-tuned on another with 5 gradient steps.}
    \label{fig:cross-problem}
\end{figure*}

\subsection{Results on Single-Problem Parameter Generation}
In this setting, we train and evaluate the neural meta-optimizer within the same problem class. 
We report the optimal hit rate, approximation ratio, and feasibility rate in Table~\ref{tab:maxcut} and~\ref{tab:mvc} for MaxCut, MIS, MaxClique, and MVC, respectively.
Additionally, we record the average number of optimization steps required for Vanilla QAOA to converge to a tolerance of $10^{-8}$.
As shown in Table~\ref{tab:maxcut} and~\ref{tab:mvc}, Uni-Meta-LSTM consistently achieves the best optimal hit rate $p(\mathbf{x}^*)$ in 14 of the 16 settings and the best approximation ratio in 12 of the 16 settings.
Furthermore, the meta-optimizer variants outperform Vanilla QAOA in several settings, particularly for constrained problem classes.
For example, on MIS at $p=8$, Uni-Meta-LSTM and Meta-LSTM achieve approximation ratios that are $3.98\%$ and $3.06\%$ higher than Vanilla QAOA, respectively, despite running only 10 optimization steps.
Additionally, our method sees a $5.65\%$ increase in optimal hit rate for the same setting.
Notably, on MaxClique, Uni-Meta-LSTM dominates the other learned baselines across all three metrics and all circuit depths, and it even surpasses Vanilla QAOA in approximation ratio at $p=6$ and in both $p(\mathbf{x}^*)$ and approximation ratio at $p=8$. 
We do, however, observe a slight performance drop from $p=8$ to $p=10$ on MIS, MaxClique and MVC, suggesting that additional model capacity or conditioning mechanisms may be needed to fully benefit from deeper circuits.
On the other hand, G2V-Meta-LSTM achieves competitive performance in a subset of settings but generally trails Meta-LSTM and Uni-Meta-LSTM. 
This suggests that purely structure-based graph embeddings may be insufficient for conditioning meta-optimizers; incorporating objective and constraint information appears important for learning transferable, instance-adaptive parameter trajectories.

\begin{figure*}[ht]
    \centering
    \includegraphics[width=\linewidth]{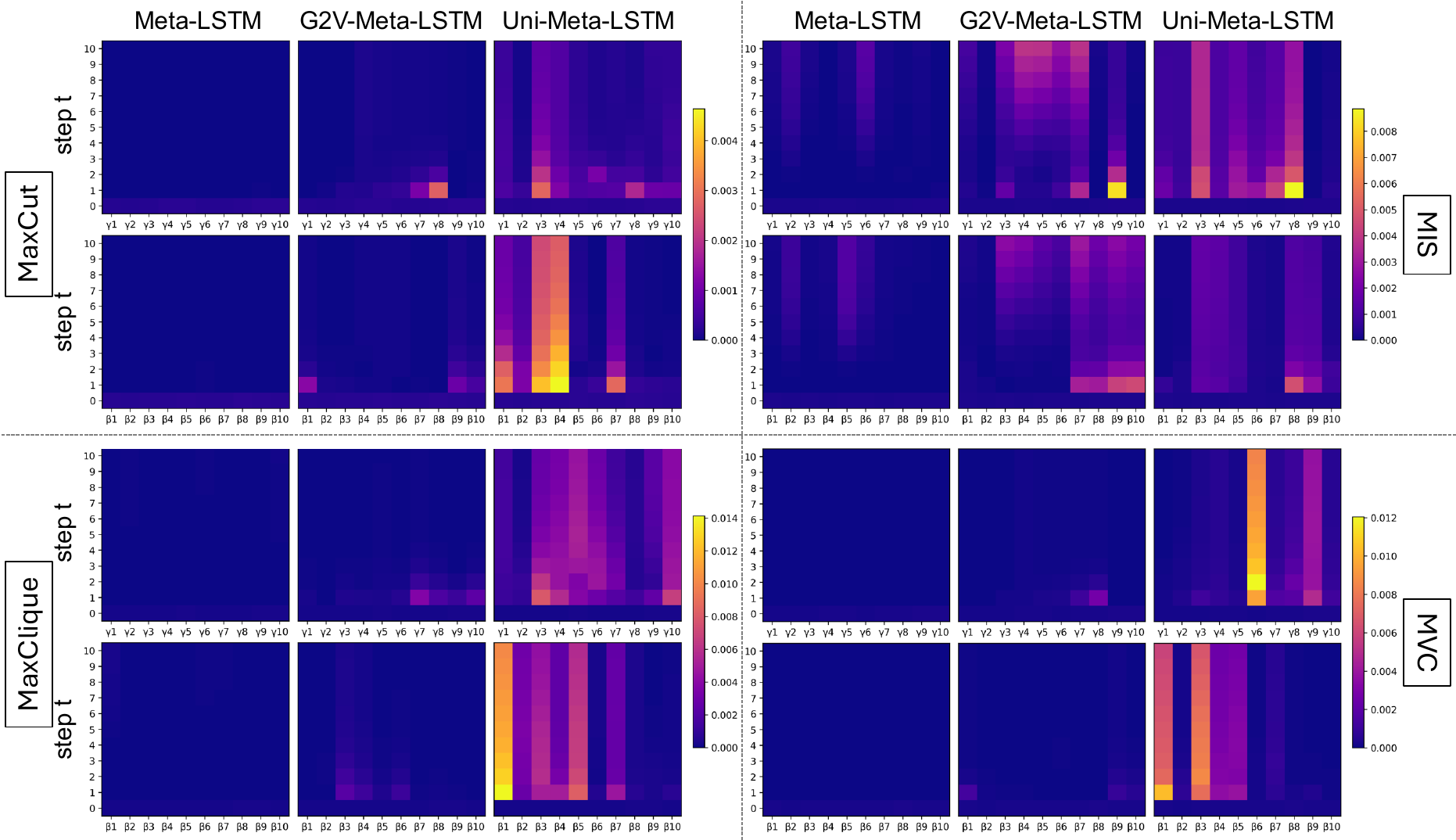}
    \caption{Visualization of diversity in generated parameter trajectories among three methods, Meta-LSTM, G2V-Meta-LSTM and Uni-Meta-LSTM, on four problem classes (MaxCut, MIS, MaxClique and MVC). The trajectories are plotted on the QAOA circuit depth $p=10$ and unrolling horizon $T=10$.}
    \label{fig:variance}
\end{figure*}

\subsection{Results on Cross-Problem Parameter Transfer}
In this experiment, we train the meta-optimizer on one problem class and directly evaluate it on a different class to assess cross-problem transferability. 
We perform exhaustive pairwise transfers among the four problem classes across all four circuit depths, yielding $4 \times 3 \times 4 = 48$ transfer settings. 
For each transfer, we additionally allow a lightweight per-instance adaptation stage in which the meta-optimizer is fine-tuned on the target instance for 5 gradient steps with a learning rate of 0.001, which is natural in our unsupervised setting since the QAOA objective provides direct training feedback without ground-truth angles.

In Figure~\ref{fig:cross-problem}, we report the optimal hit rate $p(\mathbf{x}^*)$ for the three meta-optimizer variants. 
Uni-Meta-LSTM achieves better cross-problem performance than Meta-LSTM in 34 of the 48 transfer settings, with the largest gains at depths $p\in\{4,6,8\}$ and in transfers from other problems to MaxCut and MaxClique.
G2V-Meta-LSTM occasionally improves over Meta-LSTM (e.g., in 21 transfer settings), but it still underperforms Uni-Meta-LSTM overall. 
A key limitation is that Graph2Vec provides a purely structure-based embedding; therefore, when transferring between problem classes on the same underlying graph, the conditioning signal remains essentially unchanged because the embedding does not encode the target objective or constraints. 
In contrast, Uni-Meta-LSTM conditions on problem-aware embeddings that incorporate objective and constraint information, enabling the meta-optimizer to adapt its parameter trajectory to the specific problem formulation rather than relying solely on graph structure.

\subsection{Visualization of Parameter Trajectory Expressivity}
In this section, we compare the expressivity of our graph-conditioned meta-optimizer framework with the unconditioned counterpart by visualizing the diversity in the generated parameter trajectories.
We first collect the trajectories over $N=100$ test instances in the single-problem setting.
For each instance, we run the meta-optimizer for a fixed rollout horizon $T=10$ and record the QAOA parameters generated at every step, retrieving a tensor of trajectories of dimension $N \times T \times 2p$, where $2p$ corresponds to $(\gamma,\beta)^p$.
We then quantify the trajectory diversity by computing the variance of each angle coordinate across instances at each step and aggregate these variances within $\gamma$ and $\beta$ separately.
Finally, we visualize the per-step variance patterns as heatmaps and compare the unconditioned and conditioned methods under a shared magnitude scale, where higher variance indicates more instance-dependent, thus more expressive, parameter generation.

We observe that the unconditioned Meta-LSTM tends to collapse to near-identical optimization paths, limiting its ability to generalize as problem complexity increases. In contrast, our Uni-Meta-LSTM exhibits substantially higher trajectory diversity than Meta-LSTM, indicating that incorporating instance-specific information yields more adaptive parameter generation and improved performance on more complex instances, as reflected in Tables~\ref{tab:maxcut}--\ref{tab:mvc}.



\begin{figure*}[ht]
    \centering
\includegraphics[width=0.50\linewidth]{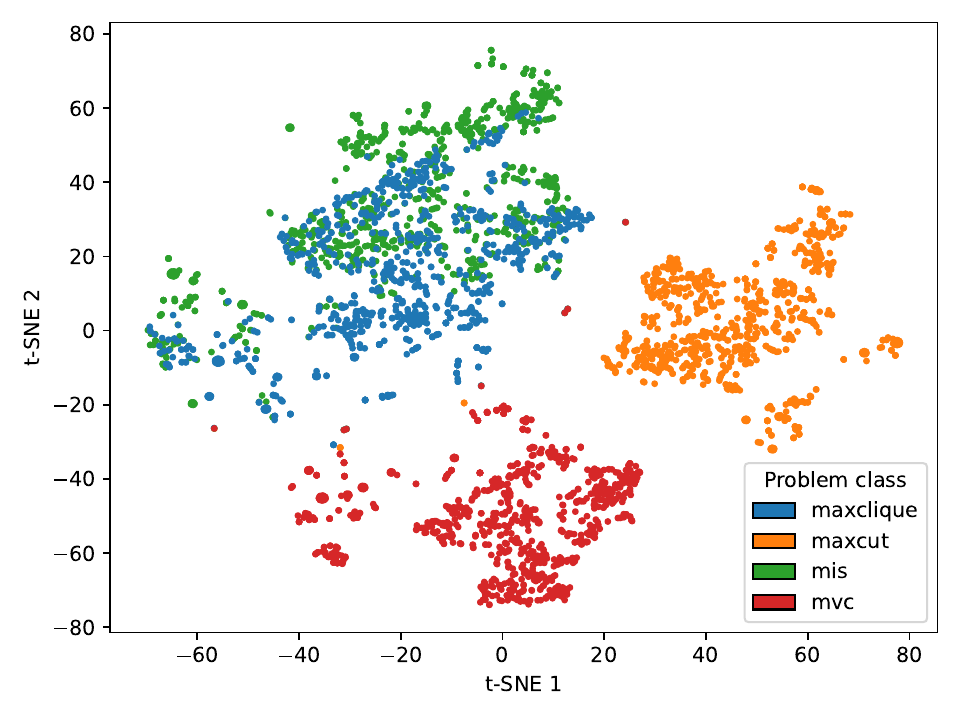}
    \caption{A demonstration of the UniHetCO embedding space: t-SNE visualization of UniHetCO graph embeddings for 1000 training instances from each of four problem classes (MaxCut, MIS, MaxClique, and MVC), totaling 4000 points. The embeddings form well-separated clusters by problem class, while also reflecting problem similarity; MIS and MaxClique exhibit partially interleaved clusters, consistent with their closely related structure.}
    \label{fig:tsne}
\end{figure*}

\subsection{Pre-training and Visualization of UniHetCO Embeddings}
In this section, we describe the UniHetCO pre-training procedure in detail. 
We then visualize the resulting problem-aware embeddings using t-distributed stochastic neighbor embedding (t-SNE)~\cite{van2008visualizing}, showing that instances form separable clusters in the embedding space. 
Finally, we discuss how the relative geometry between problem classes captures meaningful relationships that can be exploited to condition the meta-optimizer.

Given $K$ problem classes (here, $K=4$), we cast UniHetCO pre-training as a multi-domain learning problem, where each problem class defines a domain.
Let $\mathcal{L}_{\mathrm{NCO},k}$ denote the unsupervised NCO loss for class $k$. 
We train a single shared model by minimizing the mean loss across domains:
\begin{align}\label{eq:sw}
    \hat{\mathcal{L}}_{\text{NCO}}(\theta;\mathcal{D}_{\text{train}}) =  \frac{1}{K}\sum_{k=1}^K \mathcal{L}_{\text{NCO},k}(\theta; G),
\end{align}
where $G$ denotes an instance sampled from the training distribution of the corresponding domain. 
In each training iteration, we construct mini-batches by sampling an equal number of instances from each problem class, ensuring balanced exposure to all domains and preventing the training dynamics from being dominated by any single class.

After pre-training, we employ t-SNE~\cite{van2008visualizing} to visualize the relationships among the learned problem-aware graph embeddings.
We extract the embeddings from the UniHetCO trained GNN $\Phi$ on 1000 graph instances across four problem classes, resulting in 4000 points in the Euclidean space.

\noindent\textbf{Why UniHetCO Embeddings Are Informative.} As shown in Figure~\ref{fig:tsne}, the learned embeddings form well-separated clusters by problem class, indicating that the representation captures problem-specific signals beyond mere graph structure. 
Moreover, the \emph{relative geometry} between clusters is informative, as MIS and MaxClique exhibit partially interleaved regions, consistent with their closely related subset-selection structure (e.g., a clique in a graph $G$ corresponds to an independent set in its complement $\bar G$). 
This proximity provides a useful conditioning signal for the meta-optimizer, as instances that are embedded closer together are expected to induce more similar QAOA landscapes and parameter-update dynamics, thereby facilitating transfer across related formulations. 
In contrast, Graph2Vec encodes only structural information and does not incorporate the objective or constraints; consequently, the same underlying graph would receive identical embeddings regardless of the problem class, making it less suitable for guiding cross-problem instance-wise parameter generation.

\section{Conclusion and Future Work}
In this paper, we revisited the learning-to-learn paradigm for QAOA parameter generation to address an overlooked shortcoming of existing methods: \textit{near-identical parameter trajectories}. 
We thus improve instance-wise trajectory diversity by showing that graph embeddings provide an effective conditioning signal for parameter trajectory generation.
By injecting problem-aware embeddings that encode both structural and formulation-specific information, our Uni-Meta-LSTM produces more diverse and instance-adaptive parameter trajectories than unconditioned meta-optimizers, leading to improved performance on larger problem sizes and stronger cross-problem transfer. 
Results suggest incorporating objective- and constraint-aware representations is a promising direction for amortizing QAOA parameter search and improving transferability across related optimization formulations.

\noindent\textbf{Future Work.} In our empirical studies, we observe a modest performance degradation when increasing the circuit depth from $p=8$ to $p=10$, suggesting that the current conditioning mechanism may not fully exploit the additional expressivity of deeper circuits. 
Future work could investigate stronger conditioning strategies for long-horizon parameter generation, as well as training objectives that better align intermediate trajectory updates with final performance.
Beyond the current setting, an interesting direction is to train a more general meta-optimizer that can generate parameter trajectories across multiple problem classes and circuit depths within a single model, thereby enabling broader transfer and reducing the need to maintain separate models for each formulation.
\clearpage
\appendices

\section{Definition of Combinatorial Problem Classes}
We define the objectives and constraints for each problem class, followed by the conversion to the QAOA cost Hamiltonian in PennyLane~\cite{bergholm2022pennylane}.
The mixing Hamiltonian is constructed as $\sum_{i\in V} X_i$, where $X$ is the Pauli operator $X$, for all problems.
We set the initial state to $\ket{+}^{\otimes n}$.
\subsection{Maximum Cut}
The Maximum Cut problem involves finding a cut that splits a graph into two disjoint partitions so that the weighted sum of the edges within the cut is maximized.
We focus on unweighted MaxCut, setting all the edge weights to 1.
\begin{align}
\label{eq:maxcut-binary}
    \max_{\mathbf{x}\in\{0,1\}^n}\ \sum_{(u,v)\in E} x_u + x_v -2x_ux_v.
\end{align}
To construct the QAOA cost Hamiltonian, we map each binary variable to a qubit via
\begin{align}
    x_i = \frac{1 - Z_i}{2},
\end{align}
where $Z_i$ is the Pauli-$Z$ operator on qubit $i$. Substituting into~\eqref{eq:maxcut-binary} yields the cut-size operator
\begin{align}
    C(\mathbf{Z}) = \frac{1}{2}\sum_{(u,v)\in E}\big(\mathbb{I} - Z_u Z_v\big),
\end{align}
up to an additive constant. Following PennyLane's convention, we define the QAOA cost Hamiltonian as the negative of the cut-size operator:
\begin{align}
    H_C = -C(\mathbf{Z})
    = \frac{1}{2}\sum_{(u,v)\in E}\big(Z_u Z_v - \mathbb{I}\big),
    \label{eq:maxcut-hc}
\end{align}
so that minimizing $\langle H_C\rangle$ is equivalent to maximizing the MaxCut objective.

\subsection{Maximum Independent Set}
The Maximum Independent Set problem seeks the largest subset of vertices such that no two selected vertices share an edge:
\begin{align}
    \max_{\mathbf{x}\in\{0,1\}^n} \ \sum_{i\in V} x_i 
    \quad \text{s.t. } x_u + x_v \le 1,\ \forall (u,v)\in E.
\end{align}
To obtain an unconstrained formulation compatible with the standard $X$-mixer, we relax the constraints via a quadratic penalty and maximize
\begin{align}
    \max_{\mathbf{x}\in\{0,1\}^n}\ 
    \sum_{i\in V} x_i \;-\; \lambda \sum_{(u,v)\in E} x_u x_v,
    \label{eq:mis-penalty}
\end{align}
where $\lambda>0$ discourages selecting adjacent vertices. Up to an additive constant, this yields a cost Hamiltonian of the form
\begin{align}
    H_C \propto \sum_{(u,v)\in E}\big(Z_u Z_v - Z_u - Z_v\big) + \sum_{i\in V} Z_i.
\end{align}
We adopt PennyLane's unconstrained MIS convention:
\begin{align}
    H_C = 3\sum_{(u,v)\in E}\big(Z_u Z_v - Z_u - Z_v\big) + \sum_{i\in V} Z_i,
    \label{eq:mis-hc}
\end{align}
so that minimizing $\langle H_C\rangle$ promotes large independent sets while imposing a strong penalty on constraint violations.

\subsection{Maximum Clique}
The Maximum Clique problem aims to find the largest subset of vertices such that every pair of selected vertices is connected by an edge. This can be enforced by requiring that for any non-edge $(u,v)\notin E$, at most one of $u$ and $v$ can be selected:
\begin{align}
    \max_{\mathbf{x}\in\{0,1\}^n} \ \sum_{i\in V} x_i 
    \quad \text{s.t. } x_u + x_v \le 1,\ \forall (u,v)\notin E,\ u\neq v.
\end{align}
To obtain an unconstrained formulation compatible with the standard $X$-mixer, we relax the constraints via a quadratic penalty and maximize
\begin{align}
    \max_{\mathbf{x}\in\{0,1\}^N}\ 
    \sum_{i\in V} x_i - \lambda \sum_{(u,v)\in E(\bar{G})} x_u x_v,
    \label{eq:maxclique-penalty}
\end{align}
where $\lambda>0$ discourages selecting non-adjacent vertex pairs in $G$ (i.e., edges in the complement graph $\bar{G}$). Following PennyLane's unconstrained Maximum Clique convention, the QAOA cost Hamiltonian is defined as
\begin{align}
    H_C = 3\sum_{(i,j)\in E(\bar{G})}\big(Z_i Z_j - Z_i - Z_j\big) + \sum_{i\in V(G)} Z_i,
    \label{eq:maxclique-hc}
\end{align}
so that minimizing $\langle H_C\rangle$ promotes large cliques while imposing a strong penalty on violations encoded by $E(\bar{G})$.

\subsection{Minimum Vertex Cover}
The Minimum Vertex Cover problem seeks the smallest subset of vertices that covers all edges:
\begin{align}
    \min_{\mathbf{x}\in\{0,1\}^n} \ \sum_{i\in V} x_i
    \quad \text{s.t. } x_u + x_v \ge 1,\ \forall (u,v)\in E.
\end{align}
To obtain an unconstrained formulation compatible with the standard $X$-mixer, we enforce the edge-cover constraints via a penalty and minimize
\begin{align}
    \min_{\mathbf{x}\in\{0,1\}^N}\ 
    \sum_{i\in V} x_i \;+\; \lambda \sum_{(u,v)\in E} (1-x_u)(1-x_v),
    \label{eq:mvc-penalty}
\end{align}
where $(1-x_u)(1-x_v)$ equals $1$ if and only if neither endpoint of $(u,v)$ is selected, and $0$ otherwise. Following PennyLane's unconstrained MVC convention, the QAOA cost Hamiltonian is defined as
\begin{align}
    H_C = 3\sum_{(u,v)\in E}\big(Z_u Z_v + Z_u + Z_v\big) - \sum_{i\in V} Z_i,
    \label{eq:mvc-hc}
\end{align}
up to an additive constant, so that minimizing $\langle H_C\rangle$ promotes small vertex covers while penalizing uncovered edges.

\section*{Acknowledgment}
This work was supported in part by NSF award \#2444042.

\clearpage
\bibliographystyle{IEEEtran}
\bibliography{main}
\end{document}